\documentclass[showpacs,twocolumn,superscriptaddress]{revtex4-2}

\usepackage{graphicx}
\usepackage{amssymb}
\usepackage{amsmath}
\usepackage{array}
\usepackage{color}
\usepackage{hyperref} 
\hypersetup{colorlinks=true,linkcolor=blue,citecolor=cyan}
\usepackage{orcidlink}

\begin{document}

\title{Gravitational lensing of neutrinos in the Rezzolla-Zhidenko spacetime}

\author{Mirzabek Alloqulov,\orcidlink{0000-0001-5337-7117}}
\email{malloqulov@gmail.com}
\affiliation{Institute of Fundamental and Applied Research, National Research University TIIAME, Kori Niyoziy 39, Tashkent 100000, Uzbekistan}
\affiliation{University of Tashkent for Applied Sciences, Str. Gavhar 1, Tashkent 100149, Uzbekistan}
\author{Hrishikesh Chakrabarty,\orcidlink{0000-0001-5489-8846}}
\email{hrishikesh.chakrabarty@nu.edu.kz }
\affiliation{Department of Physics, Nazarbayev University, 53 Kabanbay Batyr Avenue, 010000 Astana, Kazakhstan}

\author{Daniele Malafarina,\orcidlink{0000-0002-8100-8797}}
\email{daniele.malafarina@nu.edu.kz}
\affiliation{Department of Physics, Nazarbayev University, 53 Kabanbay Batyr Avenue, 010000 Astana, Kazakhstan}
\author{Bobomurat Ahmedov,\orcidlink{0000-0002-1232-610X}}
\email{ahmedov@astrin.uz}

 \affiliation{Institute of Fundamental and Applied Research, National Research University TIIAME, Kori Niyoziy 39, Tashkent 100000, Uzbekistan}
 \affiliation{Department of Physics and Mathematics, Uzbekistan Academy of Sciences, Y. Gulomov 70, Tashkent 100047, Uzbekistan}
\author{Ahmadjon Abdujabbarov,\orcidlink{0000-0002-6686-3787}}
\email{ahmadjon@astrin.uz}

\affiliation{Institute of Fundamental and Applied Research, National Research University TIIAME, Kori Niyoziy 39, Tashkent 100000, Uzbekistan}

\date{\today}
\begin{abstract}

We consider gravitational lensing of neutrinos in the Rezzolla-Zhidenko spacetime in the weak-field limit with plane-wave approximation. 
We apply the analysis to an hypothetical system with a central object with its mass of the order of solar mass and a detector located at an Earth-like distance from the source.
We find that the deformation parameters of the Rezzolla-Zhidenko metric can have significant impact on the oscillation probabilities of the neutrinos. 
We also investigate the role of decoherence on flavor oscillations of the lensed neutrinos and show that the parameters of the Rezzolla-Zhidenko metric 
does not have significant effects on the decoherence length.
\end{abstract}

\maketitle

\section{Introduction}

Einstein's theory of general relativity (GR) has been very successful in describing strong gravitational fields and the geometric structure of spacetime and to this day it has passed all observational tests \cite{Will:2014kxa,Bambi:2015kza}. The solutions of the field equations of GR are used to describe astrophysical objects, e.g. the observable universe can be represented by the Friedmann-Robertson-Walker (FRW) metric, the exterior of a star can be approximately represented by the Schwarzschild solution, an astrophysical black hole can be represented by the Kerr solution etc. Although the most recent tests of gravity has shown excellent agreement with the theory's predictions, we are yet to explore the theory in the strong-field regime \cite{LISA:2022kgy,Cardenas-Avendano:2019zxd,Bambi:2024kqz}. One way to test the validity of the solutions of GR is to use parameterized versions of those solutions in order to obtain a number of parameters that can be constrained by observations. The goal is to compare the theoretical values of such parameters with the corresponding values obtained from existing and proposed experiments \cite{Psaltis:2009xf}. In this article we consider one of such a parameterized solutions proposed by Rezzolla and Zhidenko \cite{Rezzolla_2014} and study oscillations of neutrinos propagating in the geometry.     

The parametrization of solutions of gravitational field equations is a crucial tool for understanding the behavior of gravity at different regimes which may lead to new insights and advancements.  
This approach facilitates the comparison of different solutions, the identification of symmetries, and the study of their physical and astrophysical implications. By choosing appropriate parameters, one can simplify complex equations and gain insights into the nature of gravity, spacetime curvature, and the behavior of matter and energy within the framework of alternative gravity theories. 
It is important to note that parameterized solutions are not exact solutions of the field equations and the parameters don't always have a clear physical interpretation.
This may lead to difficulties in relating the solutions to astrophysical phenomena and measurements and therefore one needs to be careful while dealing with such solutions. 

The Rezzolla-Zhidenko (RZ) metric is a well-known spherically symmetric parameterized spacetime that deviates from Schwarzschild via a series or arbitrary parameters \cite{Rezzolla_2014}.
This metric can accommodate an infinite number of deformation parameters that determine the physical properties of the geometry at different distances from the center of symmetry.  
The RZ metric has been used to study the properties of spherically symmetric black holes and has implications for various astrophysical phenomena such as optical properties~\cite{Konoplya_2020,Suvorov_2021,Konoplya_2022,Konoplya_2023,Broderick_2023,Shaymatov_2023}, gravitational lensing~\cite{Kocherlakota_2020}, accretion disks~\cite{Kocherlakota_2022,Bauer_2022}, gravitational waves~\cite{V_lkel_2020,C_rdenas_Avenda_o_2020} etc. 

Assuming that the RZ can be used to model the exterior of an astrophysical object, such as a stellar mass black hole, our aim is to study the propagation of neutrinos and neutrino oscillations in the geometry.
Neutrino oscillations refer to the phenomenon where neutrinos, which are elementary particles which weakly interact with matter, change their flavor as they propagate through regions of spacetime. The phenomenon of neutrino oscillation is affected by the presence of a gravitational field, i.e. in curved spacetime.
This means that the equations governing neutrino propagation depend on the geometry, thus leading to changes in the oscillation behavior of neutrinos through changes in the trajectory and energy of neutrinos, which could in principle be detected. 
Consequently, the probabilities of neutrinos transitioning between different flavors, such as electron neutrinos, muon neutrinos, and tau neutrinos, can be altered as compared to the flat spacetime scenario. This has led to several theoretical studies on neutrino oscillation in curved spacetime \cite{Cardall:1996cd,Piriz:1996mu,Ahluwalia:1996ev,Bhattacharya:1999na,Pereira:2000kq,Crocker:2003cw,Lambiase:2005gt,Godunov:2009ce,Ren:2010yf,Geralico:2012zt,Chakraborty:2013ywa,Visinelli:2014xsa,Zhang:2016deq,Alexandre:2018crg,Blasone:2019jtj,Buoninfante:2019der,Boshkayev:2020igc,Mandal:2021dxk,Koutsoumbas:2019fkn,Wudka:1991tg,Fornengo_1997,Swami_2020,Swami_2021,Chakrabarty:2021bpr,Chakrabarty:2023kld}. 

Since the geometry plays a crucial role in the propagation and oscillations of neutrinos, it is important to estimate the effects through a parametrized solution such as the RZ metric. 
The study of neutrino oscillations in curved spacetime is crucial for understanding the behavior of neutrinos in astrophysical environments, such as near supernovae or in the vicinity of massive compact objects. The hope is that by investigating neutrino oscillations in curved spacetime we may gain insights into the fundamental properties of neutrinos and their interactions with gravity \cite{Swami_2020,Chakrabarty:2021bpr}.
Our goal in this paper is to study how the parametrized deviations from the Schwarzschild geometry affect the flavor transition probabilities of neutrinos. We start with treating the neutrinos in plane-wave approximation and derive the phase of oscillation which is used to determine the probabilities of flavor transition. Later we treat the neutrinos in the wave-packet approximation to understand the decoherence effects in RZ spacetime. 

The article is organized as follows: In sections \ref{sec-2} and \ref{sec-3}, we discuss the RZ spacetime metric and the basics of neutrino oscillations in flat spacetime. In Section \ref{sec-4}, we derive the phase of oscillation in RZ spacetime while section \ref{sec-5} is devoted to the numerical study of lensing effects on the neutrino flavor transition. In Section \ref{sec-6}, we discuss the decoherence properties. Finally in Section \ref{sec-7} we summarize the results and discuss future prospects.  
Throughout the article we use units for which $c = G = 1$ and employ the $(-,+,+,+)$ metric signature.

\section{Rezzolla-Zhidenko metric}\label{sec-2}

\begin{figure*}
    \centering
    \includegraphics[scale=0.6]{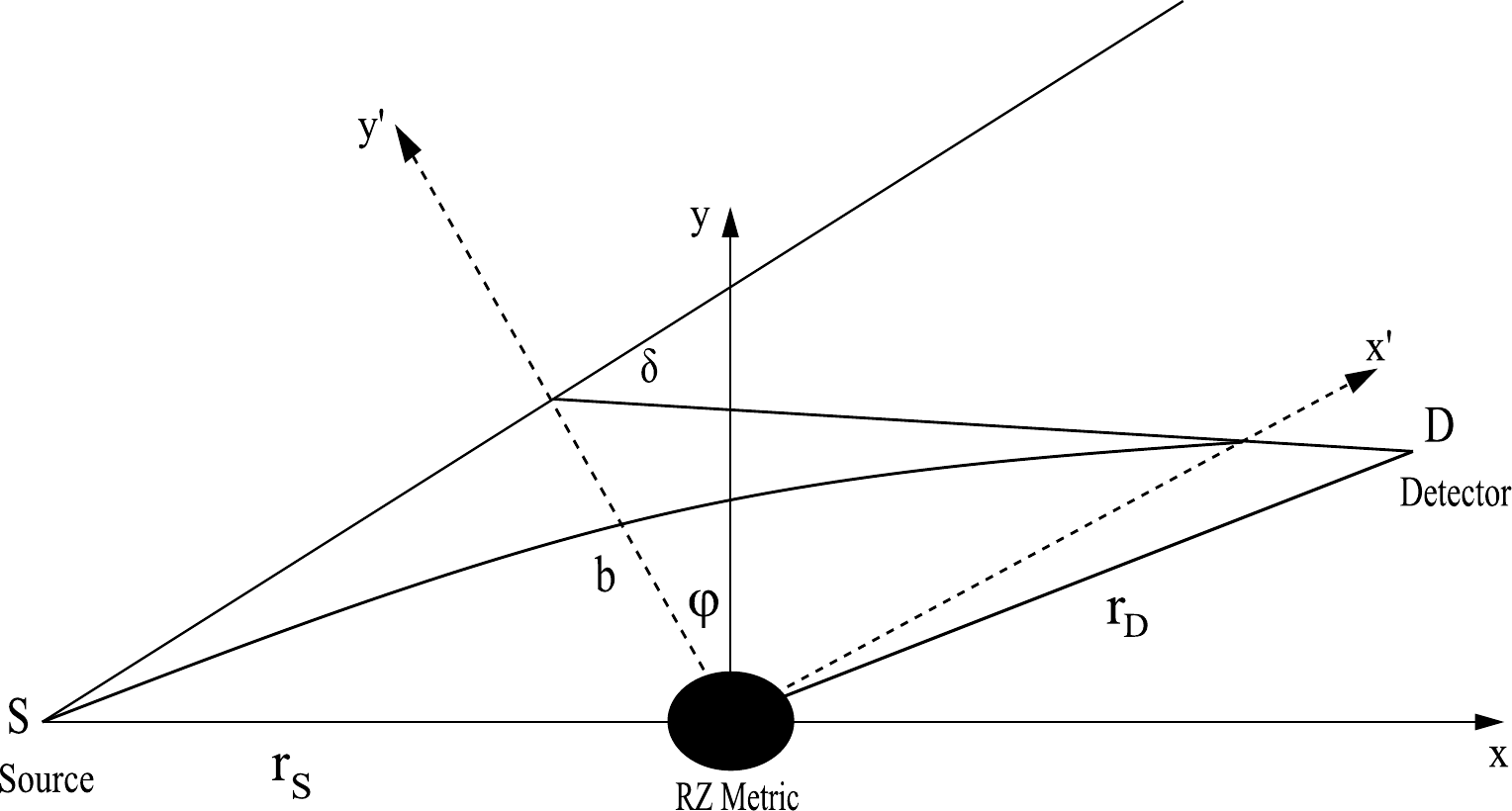}
    \caption{Schematic diagram of weak lensing of neutrinos in the RZ spacetime. Neutrinos propagate from the source $S$ to detector $D$ in the exterior of a static and spherical massive object described by the RZ spacetime.}
    \label{Fig:weak lensing}
\end{figure*}

The line element of a general spherically symmetric static metric in a spherical coordinates $\{t,r,\theta,\phi\}$ can be written as \cite{Rezzolla_2014}
\begin{equation}\label{line element}
    ds^2=-N^2(r)dt^2+\dfrac{B^2(r)}{N^2(r)}dr^2+r^2 d \Omega^2,
\end{equation}
where $d \Omega^2=d \theta^2+\sin^2{\theta} d \phi^2$ is the usual spherical part of the metric, and $N$ and $B$ are functions of the radial coordinate $r$ only. The radial location of the event horizon is marked as $r = r_0 > 0$ and this definition implies that 
\begin{equation}
   N^2(r_0)=0.
\end{equation}
We neglect any cosmological effect, so that the line element \eqref{line element} can be taken as asymptotically flat. Then the radial coordinate may be compactified by introducing a dimensionless variable $x$, given by
\begin{equation}
    x=1-\dfrac{r_0}{r}.
\end{equation}
Evidently, $x=0$ corresponds to the location of the event horizon and $x=1$ corresponds to spatial infinity. The metric function $N(r)$ can be rewritten in terms of the dimensionless variable $x$ as
\begin{equation}\label{eq:n}
    N^2(x)=x A(x),
\end{equation}
with 
\begin{equation}\label{eq:A}
    A(x)>0 \hspace{0.5cm} \text{for} \hspace{0.5cm} 0\leq x \leq 1.
\end{equation}
Now the functions A and B  can be rewritten by introducing dimensionless parameters $\epsilon$, $a_0$ and $b_0$ as 
\begin{eqnarray}
   && A(x)=1-\epsilon(1-x)+a_0(1-x)^2+\Tilde{A}(x)(1-x)^3, \\
   && B(x)=1+b_0(1-x)+\Tilde{B}(x)(1-x)^2,
\end{eqnarray}
where the functions $\Tilde{A}(x)$ and $\Tilde{B}(x)$ are used to describe the asymptotic behavior of the metric. 

In \cite{Rezzolla_2014}, the authors introduce Padé series and express the functions $\Tilde{A}(x)$ and $\Tilde{B}(x)$ in terms of continued fractions
\begin{subequations}
    \begin{equation}
        \Tilde{A}(x)=\dfrac{a_1}{1+\dfrac{a_2 x}{1+\dfrac{a_3 x}{1+ ...}}},
    \end{equation}
     \begin{equation}
        \Tilde{B}(x)=\dfrac{b_1}{1+\dfrac{b_2 x}{1+\dfrac{b_3 x}{1+ ...}}},
    \end{equation}
\end{subequations}
where  $a_1, a_2 , a_3 ... $ and  $b_1, b_2 , b_3 ... $ are dimensionless constants.
Then the metric functions in (\ref{line element}) can be expanded up to any desired order. We obtain
\begin{eqnarray}
   N^2(x)&=&1-(1+\epsilon)(1-x)+a_0(1-x)^2+ \\ \nonumber
  && +(a_1-a_0+\epsilon)(1-x)^3 -a_1(1-x)^4, \\
    B^2(x)&=&\left(1+ b_0 (1-x)+b_1(1-x)^2\right)^2.
\end{eqnarray}
%
Notice here that the parameter $\epsilon$ describes the departure of the event horizon radius $r_0$ from $2M$ as it is related to the event horizon by
%
\begin{equation}\label{epsilon}
    \epsilon=-\left(1-\dfrac{2 M}{r_0}\right),
\end{equation}
where $M$ is the ADM mass of the spacetime. In order to retrieve the Schwarzschild spacetime for which the horizon is at $r_0=2M$ we must have $\epsilon=0$ and $a_i=b_i=0$ ($i=0,1,2...$). 
In general, the functions $N$ and $B$ can be expressed in terms of the parameterized post-Newtonian (PPN) parameters as \cite{Konoplya:2020hyk}

\begin{eqnarray}
   N^2&=&1-\frac{2M}{r}+(\beta-\gamma)\frac{2M^2}{r^2}+o(1/r^3), \\
    \frac{B^2}{N^2}&=& 1 + \gamma \frac{2M}{r}+o(1/r^2),
\end{eqnarray}
so that we get
\begin{eqnarray} \label{beta}
   \beta - \gamma &=& \frac{2a_0}{(1+\epsilon)^2}, \\ \label{gamma}
    \gamma -1 &=& \frac{2b_0}{1+\epsilon}.
\end{eqnarray}

The parametrization then allows to describe deviations from the Schwarzschild metric at every perturbative order. In the following we shall return to spherical coordinates and write the line element Eq.~(\ref{line element}) in the following form
\begin{equation}
    d s^2=-\mathcal{A}dt^2+\mathcal{B}dr^2+\mathcal{C}d \theta^2+\mathcal{D}d \phi^2
\end{equation}
where $\mathcal{A}$ and $\mathcal{B}$ are defined from $A$ and $B$ by keeping only first order terms, namely $\epsilon\neq 0$ and $b_0\neq 0$ with all the other parameters vanishing. This gives
\begin{equation}
    \begin{aligned}
        \mathcal{A} (r) &= N^2(r) = 1-(1+\epsilon)\Big(\frac{r_0}{r}\Big), \\
        \mathcal{B} (r) &= \dfrac{B^2(r)}{N^2(r)} = \frac{\left(1+b_0 \Big(\frac{r_0}{r}\Big)\right)^2}{1-(1+\epsilon)\Big(\frac{r_0}{r}\Big)}, \\ 
        \mathcal{C} (r) &=r^2, \\
        \mathcal{D} (r) &=r^2 \sin^2{\theta}.
    \end{aligned}
\end{equation}

\section{Neutrino oscillations in flat spacetime}\label{sec-3}

\begin{figure*}
    \includegraphics[scale=0.4]{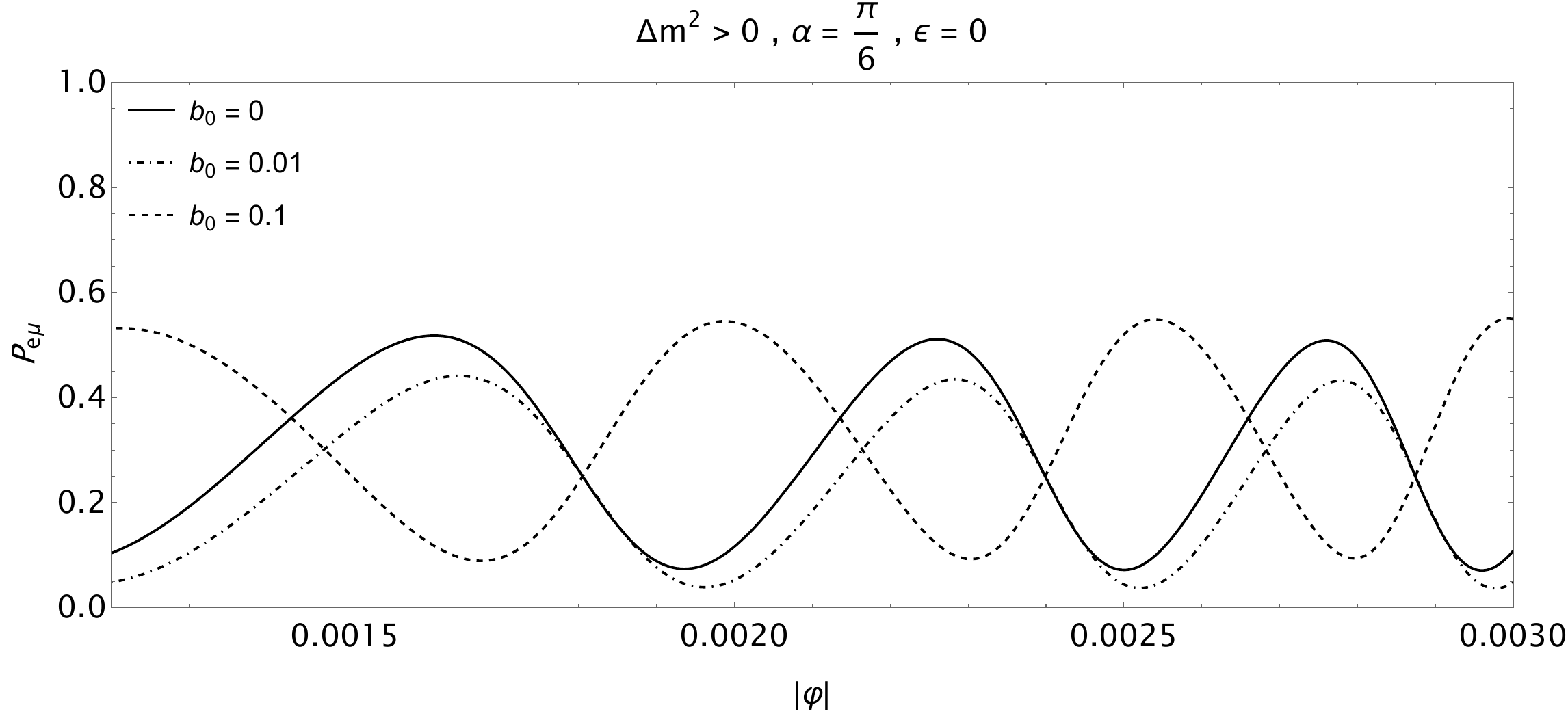}
    \includegraphics[scale=0.4]{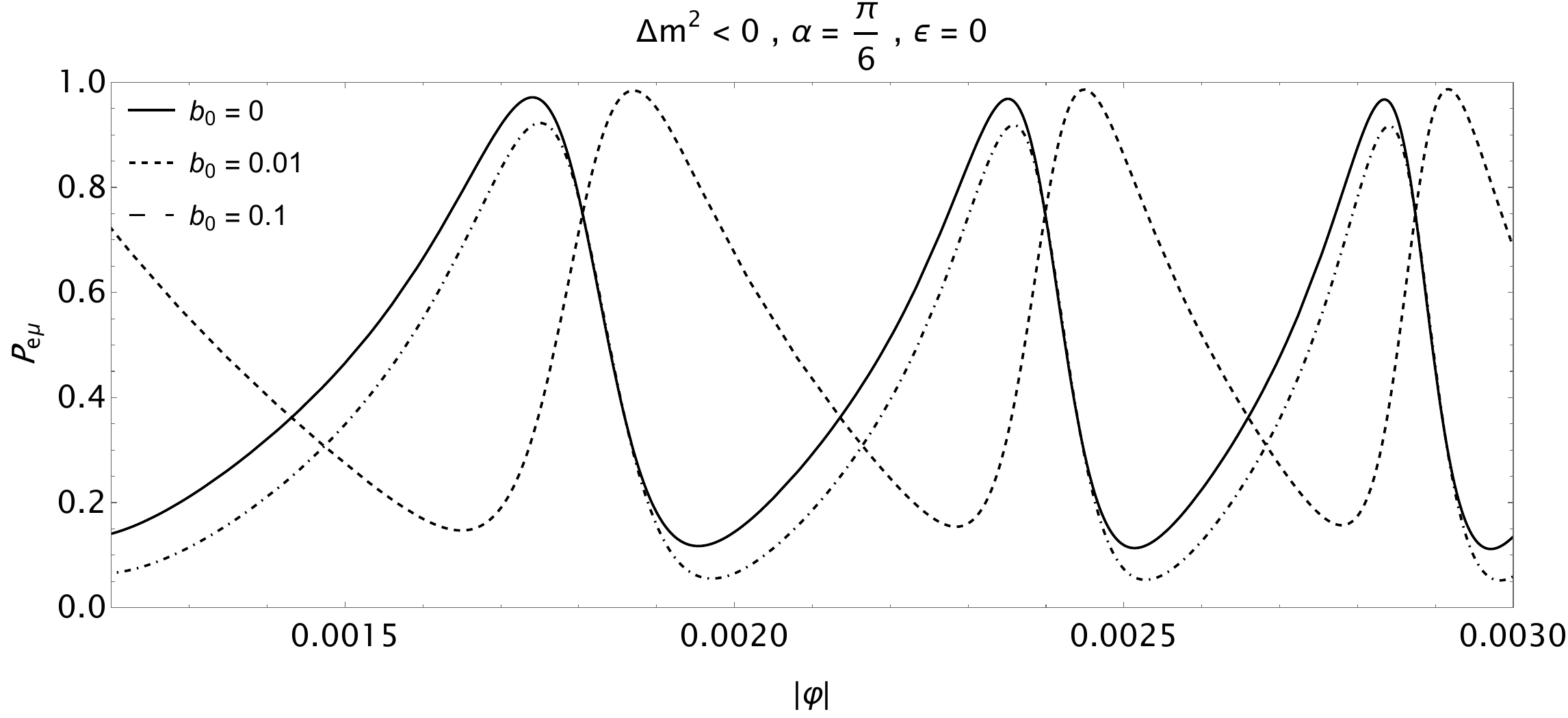}
    \caption{Top panel: neutrino oscillation probability as a function of azimuthal angle $\varphi$ for $b_0 = 0$ (solid line), $b_0 = 0.01$ (dotted-dashed line) and $b_0 = 0.1$ (dashed line) for normal hierarchy $\Delta m^2 > 0$.  Bottom panel: neutrino oscillation probability for $b_0 = 0$ (solid line), $b_0 = 0.01$ (dotted-dashed line) and $b_0 = 0.1$ (dashed line) for inverted hierarchy $\Delta m^2 < 0$. The mixing angle here is $\alpha = \pi/6$. Values of the other parameters are as follows: $M = 1M_{\odot}$, $\Delta m^2=10^{-3}$ $eV^2$, and the lightest neutrino here is considered to be massless.}
\label{fig2}
\end{figure*}

In weak interactions, neutrinos are produced and detected in different flavor eigenstates denoted by $|\nu_{\alpha}\rangle$ with $\alpha$ =$e$, $\mu$, $\tau$, and the flavor eigenstates are superposition of mass eigenstates denoted by $\nu_i$ where $i=1, 2, 3$. One can write the relations between mass and flavor eigenstates
\begin{equation}
    |\nu_{\alpha}\rangle=\sum_i U^{*}_{\alpha i}|\nu_{i}\rangle,
\end{equation}
where $U$ is the $3 \times 3$ unitary mixing matrix. For three flavor neutrino oscillation, this is known as
the Pontecorvo–Maki–Nakagawa–Sakata (PMNS) leptonic
mixing matrix~\cite{Pontecorvo:1957qd,10.1143/PTP.28.870,Pontecorvo:1967fh}.
We assume that the neutrino wave-function is a plane wave
as considered originally in~\cite{Pontecorvo:1957qd,10.1143/PTP.28.870,Pontecorvo:1967fh} and it propagates from a
source $S$ located at a spacetime event ($t_S$,$x_S$) to a detector $D$
located at a spacetime event ($t_D$,$x_D$). So the wave-function at the detector point is given by
\begin{equation}
   |\nu_i(t_D,x_D)\rangle=\exp{(-i \Phi_i)}|\nu_i(t_S,x_S)\rangle,
\end{equation}
where $\Phi_i$ is the phase of oscillation. Neutrinos are expected to be produced initially in the flavour eigenstate $|\nu_{\alpha}\rangle$ at $S$ and then travel to the detector $D$.
In that case, the probability of the change in neutrino flavour
from $\nu_{\alpha}$ to $\nu_{\beta}$ at $D$ is given by 
\begin{eqnarray}
    {\cal P}_{\alpha \beta}&=&|\langle\nu_{\beta}|\nu_{\alpha}(t_D,x_D)\rangle |^2 = \nonumber \\
 &=&\sum_{i,j}U_{\beta i}U^{*}_{\beta j}U_{\alpha j}U^{*}_{\alpha i}\exp{(-i(\Phi_i-\Phi_j))}.
\end{eqnarray}
The change in flavor can occur if $\Phi_i \not = \Phi_j$. Different neutrino
mass eigenstates develop different phases $\Phi_i$ because of differences in their mass and energy/momentum which ultimately gives rise to neutrino oscillation phenomena~\cite{Akhmedov2009}. 
In flat spacetime, the phase is given by 
\begin{equation}\label{phase}
\Phi_i=E_i(t_D-t_S)-{\bf p}_i \cdot ({\bf x}_D-{\bf x}_S).
\end{equation}
It is typically assumed that all the mass eigenstates in a flavor eigenstate initially produced at the source have equal
momentum or energy~\cite{Akhmedov2009,Akhmedov2011}. Either of these assumptions together with $(t_D-t_S) \simeq |{\bf x_D-x_S}|$ for relativistic neutrinos ($E_i\gg m_i$) leads to
\begin{equation}
    \Delta \Phi_{i j}=\Phi_i-\Phi_j \simeq \frac{\Delta m^2_{i j}}{2 E_0}|x_D-x_S|,
\end{equation}
where $\Delta m^2_{ij}=m^2_i-m^2_j$, and $E_0$ is the average energy of the relativistic neutrinos produced at the source. 
To generalize the expression of the phase $\Phi_i$ for neutrino propagation in curved spacetime, Eq.(\ref{phase}) is written in the
covariant form
\begin{equation}
    \Phi_i= \int_S^D p^{(i)}_{\mu} d x^{\mu},
\end{equation}
where
\begin{equation}\label{p(i)}
    p^{(i)}_{\mu}=m_i g_{\mu \nu} \frac{d x^{\nu}}{d s},
\end{equation}
is the canonical conjugate momentum to the coordinates $x^{\nu}$
and $g_{\mu \nu}$ and $d s$ are the metric tensor and line element of the
curved spacetime, respectively.

\section{Phase of lensed neutrinos}\label{sec-4}

\begin{figure*}
    \includegraphics[scale=0.4]{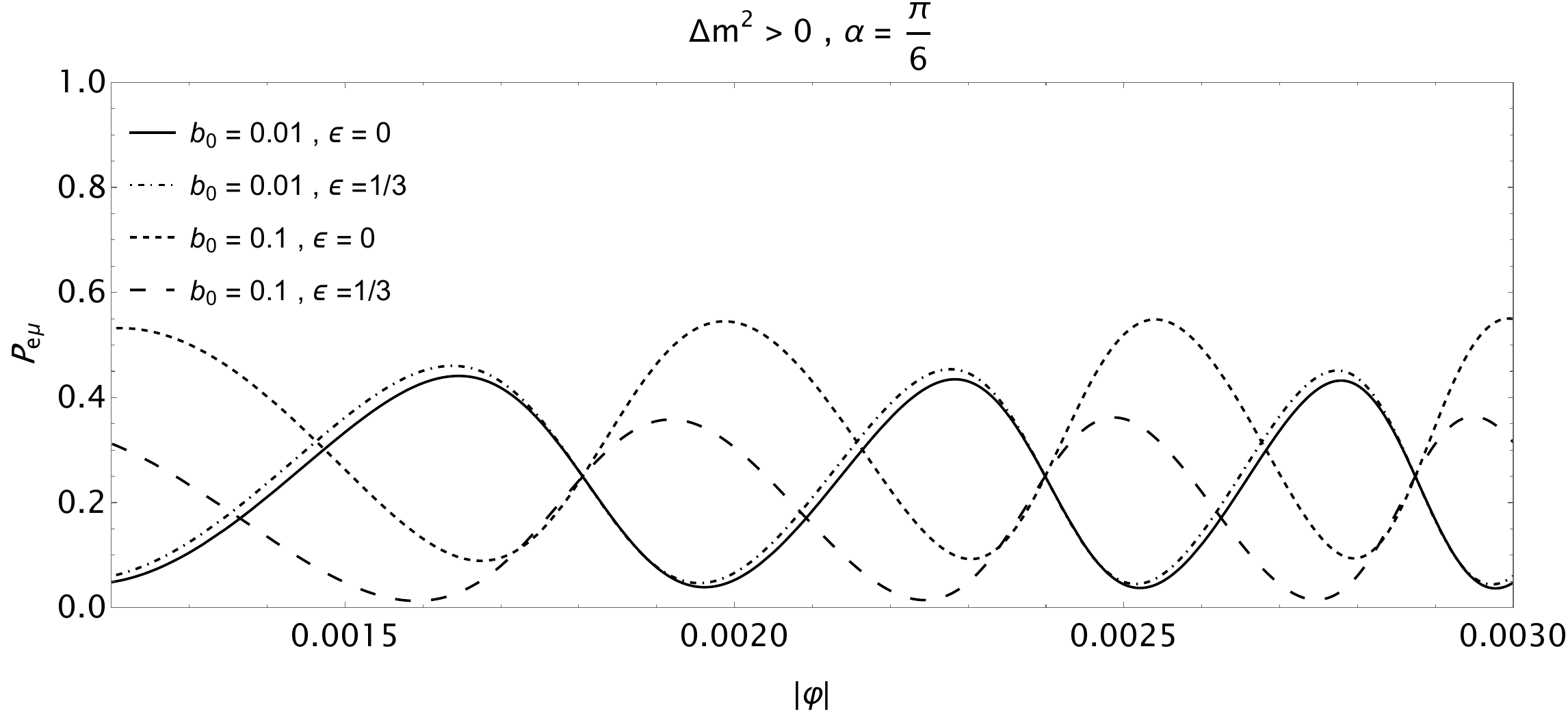}
    \includegraphics[scale=0.4]{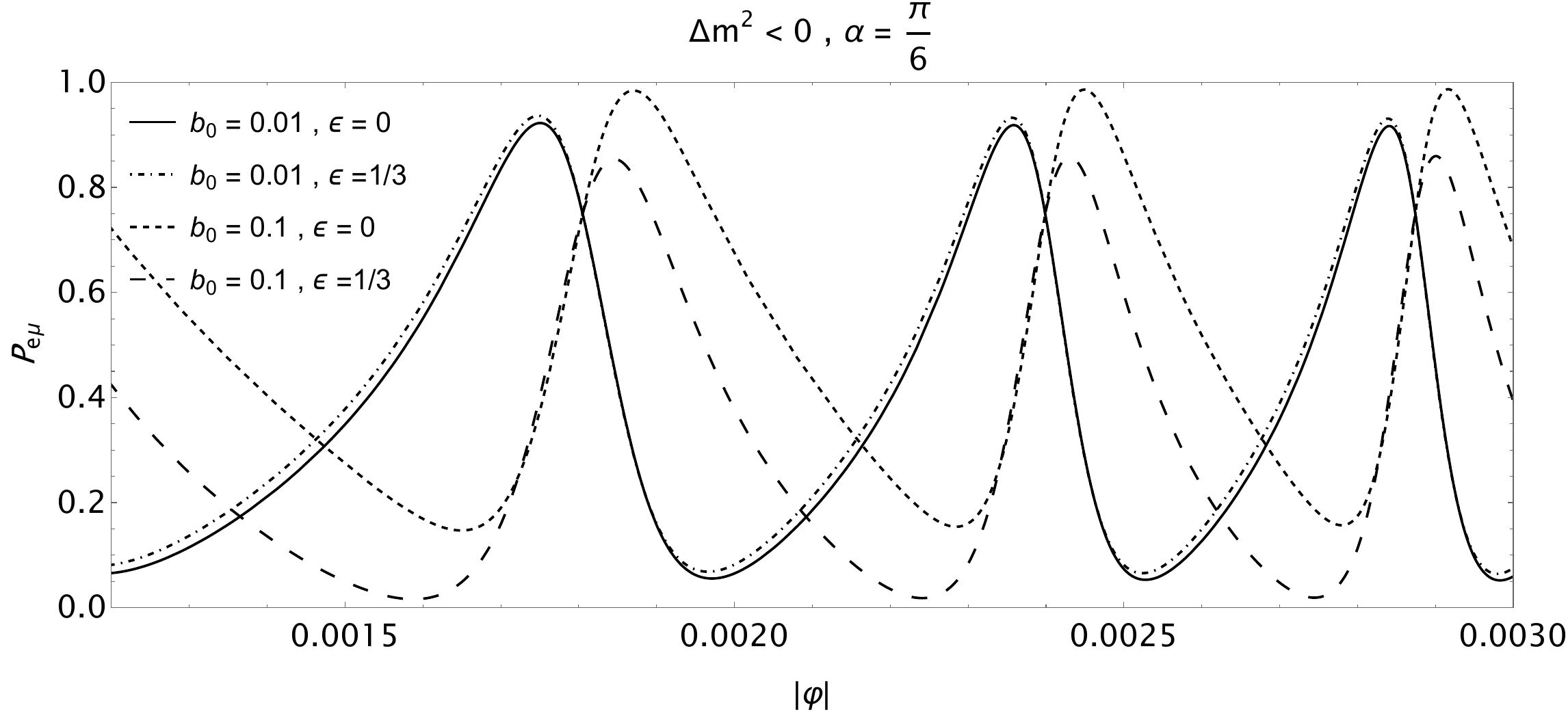}
   \caption{Top panel: neutrino oscillation probability as a function of azimuthal angle $\varphi$ for $b_0 = 0.01$, $\epsilon=0$ (solid line), $b_0 = 0.01$, $\epsilon=1/3$ (dotted-dashed line), $b_0=0.1$, $\epsilon=0$ (dashed line) and $b_0 = 0.1$, $\epsilon=1/3$ (large dashed line) for normal hierarchy $\Delta m^2 > 0$.  Bottom panel: neutrino oscillation probability for $b_0 = 0.01$, $\epsilon=0$ (solid line), $b_0 = 0.01$, $\epsilon=1/3$ (dotted-dashed line), $b_0=0.1$, $\epsilon=0$ (dashed line) and $b_0 = 0.1$, $\epsilon=1/3$ (large dashed line) for inverted hierarchy $\Delta m^2 < 0$. The mixing angle here is $\alpha = \pi/6$. Values of the other parameters are as follows: $M = 1M_{\odot}$, $\Delta m^2=10^{-3}$ $eV^2$.}
\label{fig3}
\end{figure*}

To evaluate the phase of neutrino oscillation probability we first define the components of canonical momenta $p_{\mu}^{(k)}$ (where the superscript $k$ refers to the neutrino mass eigenstate) for test particles moving in the equatorial plane $\theta=\pi/2$ from Eq.\eqref{p(i)} with $p_{\theta}^{(k)}=0$. The metric components do not depend on the coordinates $t$ and $\phi$ ensuring that their canonical momenta $p_t^{(k)}$ and $p_{\phi}^{(k)}$ are constant along the trajectory of particles. We can define the constants of motion as $p_t^{(k)}=-E_k$ and $p_{\phi}^{(k)}=J_k$. This allows us to drop the subscript $r$ from the momentum $p_r^{(k)}$ to ease the notation and define $p_r^{(k)}=p_k$.
The mass of the k-th eigenstate $m_k$ can then be written using the mass-shell relation as  
\begin{equation}\label{Eq:mass-shell}
    -m_k^2=g^{tt} E_k^2+g^{rr} p_k^2+g^{\phi \phi} J_k^2.
\end{equation}
We can now look at the oscillation phase for both radial and non-radial propagation of neutrinos.  

\subsection{Radial propagation}

First we investigate the radial propagation of neutrinos in the equatorial plane thus setting $J_k =0$ and $ \dot{\theta} =0$. From Eq.\eqref{p(i)}, we have 
\begin{equation}\label{relations}
    \dfrac{d t }{d s}=-\dfrac{E_k}{m_k g_{tt}}, \hspace{0.5cm} \dfrac{d r }{d s}=\dfrac{p_k}{m_k g_{rr}}.
\end{equation}
Using these definitions, the phase of a neutrino oscillation can be written as
\begin{equation}
    \Phi_k= \int_S^D \left[-E_k \left(\dfrac{d t}{d r}\right)_0+p_k(r)\right]dr,
\end{equation}
where, $S$ and $D$ denote the source and detector of the neutrinos, respectively and the subscript $0$ refers to a light-ray trajectory. From Eq.\eqref{relations}, we can write the light-ray differential as
\begin{equation}\label{differential}
    \left(\dfrac{dt}{dr}\right)_0=\dfrac{E_0}{p_0(r)}\dfrac{\mathcal{B}}{\mathcal{A}},
\end{equation}
where $E_0$ and $ p_0(r) $ are the energy and momentum of a massless particle at infinity. Then the mass-shell relation can be written in a unified way for massive and massless particles as
\begin{eqnarray}\label{momenta}
   && p_k(r)=\pm \sqrt{\dfrac{\mathcal{B} E_k^2}{\mathcal{A}}-\mathcal{B} m_k^2},
\end{eqnarray}
where the massless case is obtained for $k=0$ and $m_0=0$.
Now using Eq.\eqref{differential} and \eqref{momenta} we can rewrite the phase as
\begin{equation}
     \Phi_k= \pm \int_S^D E_k \sqrt{\dfrac{\mathcal{B}}{\mathcal{A}}}\left[-1+\sqrt{1-\dfrac{m_k^2 \mathcal{A}}{E_k^2}}\right]dr.
\end{equation}
Using Eqs.\eqref{eq:n} and \eqref{eq:A} we expand square root under the bracket of the above expression to obtain
\begin{equation}
     \Phi_k= \pm \int_S^D \sqrt{\mathcal{A}\mathcal{B}} E_k \dfrac{m_k^2}{2 E_k^2} dr,
\end{equation}
which further simplifies the analysis. Now we use relativistic approximation ($m_k<<E_k$), following \cite{Fornengo_1997} 
\begin{eqnarray}
  && E_k \simeq E_0 + \mathcal{O}\left(\dfrac{m_k^2}{2 E_0}\right), \nonumber \\
  && E_k \dfrac{m_k^2}{2 E_k^2} \simeq E_0 \dfrac{m_k^2}{2 E_0^2},
\end{eqnarray}
and the phase of oscillation for radially propagating neutrinos in the equatorial plane becomes 
\begin{equation}
  \Phi_k= \pm \dfrac{m_k^2}{2 E_0}\int_S^D  \sqrt{\mathcal{A} \mathcal{B}} d r.
\end{equation}
In the Schwarzschild case we have that $\mathcal{A}\mathcal{B}=1$ and the integral simply gives $r_D-r_S$. On the other hand, for the RZ metric there will be additional terms depending on the expansion parameters.
In fact, integrating the above integral from $r_S$ to $r_D$, we get

\begin{equation}\label{eq:r1}
    \Phi_k \approx \pm \dfrac{m_k^2}{2 E_0}\left[r_D-r_S+\dfrac{2 b_0 M}{1+\epsilon}  \ln{\dfrac{r_D}{r_S}}\right] .
\end{equation}

Notice that the phase does not depend on $a_0$ at this order, since $a_0$ is related to the term in $M^2/r^2$ in the PPN expansion. At the same time the dependence on $\epsilon$ at large distances is subdominant with respect to $b_0$ and it disappears once we measure the phase using the PPN parameter $\gamma$ as in Eq.~\eqref{gamma}. Notice also that for $b_0\rightarrow0$, we can recover the result of the phase for a radially propagating neutrinos in the Schwarzschild spacetime as~\cite{Fornengo_1997,Swami_2020}, namely
\begin{equation}\label{Schw}
    \Phi_k \approx \pm \dfrac{m_k^2}{2 E_0}|r_D-r_S|.
\end{equation}

\subsection{Non–radial propagation}

We turn now the attention to the case of non-radial propagation of neutrinos in the equatorial plane. In this case $J_k\neq 0$ and the phase is given by 
\begin{equation}
    \Phi_k=\int_S^D \left[-E_k\left(\dfrac{d t}{d r}\right)_0 + p_k + J_k \left(\dfrac{d \phi}{d r}\right)_0\right] d r,
\end{equation}
where $J_k$ is the angular momentum of the $k$-th mass eigenstate of the neutrino. The integral is again taken along the light–ray trajectory that links the source $S$ to the detector $D$. Now from Eq.\eqref{momenta} we can obtain the following relations
\begin{equation}
    \dfrac{d t }{d r }=\dfrac{E_k}{p_k}\dfrac{\mathcal{B}}{\mathcal{A}} , \hspace{0.5cm} \dfrac{d \phi }{d r }=\dfrac{J_k}{p_k}\dfrac{\mathcal{B}}{\mathcal{D}}, 
\end{equation}
which along the light-ray trajectories become
\begin{equation}\label{Eq:light-ray}
    \left(\dfrac{dt}{dr}\right)_0=\dfrac{E_0}{p_0}\dfrac{\mathcal{B}}{\mathcal{A}} ,\hspace{0.5cm}  \left(\dfrac{d\phi}{dr}\right)_0=\dfrac{J_0}{p_0}\dfrac{\mathcal{B}}{\mathcal{D}}.
\end{equation}
For convenience, we express $J_k$ as a function of the energy $E_k$ as
\begin{equation}\label{Eq:angular}
    J_k=E_k b v_k^{\infty},
\end{equation}
where $b$ and $v_k^{\infty}$ are the impact parameter and the velocity at infinity respectively. Since the metric is asymptotically flat, we can write 
\begin{eqnarray}\label{Eq:vinfty}
    && v_k^{\infty}=\dfrac{\sqrt{E_k^2-m_k^2}}{E_k} \simeq 1-\dfrac{m_k^2}{2 E_k^2}, \nonumber \\
    && J_k \simeq E_k b \left(1-\dfrac{m_k^2}{2 E_k^2}\right),
\end{eqnarray}
where in the second equality we used the relativistic approximation up to the order $\mathcal{O}$($m_k^2/E_k^2$). The angular momentum of a massless particle now is
\begin{equation}\label{Eq:massless}
    J_0=E_0 b,
\end{equation}
and using Eqs.(\ref{Eq:light-ray})-(\ref{Eq:massless}), the phase becomes
\begin{equation}
    \Phi_k = \int _S^D \dfrac{E_0 E_k \mathcal{B}}{p_0}\left[-\dfrac{1}{\mathcal{A}}+\dfrac{p_0 p_k}{E_0 E_k \mathcal{B}}+\dfrac{b^2}{\mathcal{D}}\left(1-\dfrac{m_k^2}{2 E_k^2}\right)\right]d r.
\end{equation}
Now we shall use the mass-shell relation to bring the phase integral to a convenient form for integration. We use Eq.\eqref{Eq:mass-shell} to write
\begin{equation}\label{Eq:relations}
    \dfrac{p_0(r) p_k(r)}{E_0 E_k \mathcal{B}}=\dfrac{1}{\mathcal{A}}-\dfrac{b^2}{\mathcal{D}}-\dfrac{m_k^2}{2 E_k^2},
\end{equation}
which is valid also in the massless case by setting $k=0$ with $m_0=0$. Using the above expressions we write the phase as
\begin{equation}
    \Phi_k = - \int_S^D \dfrac{E_0 \mathcal{B}}{p_0(r)} E_k \dfrac{m_k^2}{2 E_k^2} d r .
\end{equation}
Finally we again use the relativistic approximation to get
%
\begin{eqnarray} \nonumber
        \Phi_k &=&-\dfrac{m_k^2}{2 E_0} \int_S^D \dfrac{E_0 \mathcal{B}}{p_0}d r = \\
        &=& \pm \dfrac{m_k^2}{2 E_0} \int_S^D \sqrt{\mathcal{A} \mathcal{B}}\left(1-\dfrac{b^2 \mathcal{A}}{\mathcal{D}}\right)^{-1/2}d r.
    \label{Eq:phase2}
\end{eqnarray}
The above integral represents the phase of oscillation of neutrinos travelling non-radially in the equatorial plane of the RZ spacetime. We can now divide the analysis of the integral in two cases. First, we shall consider the case where a neutrino escapes the gravitational potential of the RZ metric and then propagates outwards non–radially. We use the weak-field approximation and expand the integrand in Eq.(\ref{Eq:phase2}) as
\begin{eqnarray}\nonumber
        \Phi_k &\approx& \pm \dfrac{m_k^2}{2 E_0} \int_S^D \left[\dfrac{1}{\sqrt{1-\dfrac{b^2}{r^2}}}+\right. \\
        &&+\left.\dfrac{[2 b_0 (r^2-b^2)/(1+\epsilon) - b^2]M/r}{(r^2-b^2){\sqrt{1-\dfrac{b^2}{r^2}}}}\right]d r.
\end{eqnarray}
This can be easily integrated analytically from the location of the source $ r_S $ to the detector $ r_D $. We get

\begin{eqnarray}
\Phi_k &\approx& \dfrac{m_k^2}{2 E_0} \left[\sqrt{r_D^2-b^2}-\sqrt{r_S^2-b^2}+M\left(\dfrac{r_D}{\sqrt{r_D^2-b^2}}+ \right. \right. \nonumber \\
&&  -\left.\left.\dfrac{r_S}{\sqrt{r_S^2-b^2}}\right)+\dfrac{2 b_0 M}{ 1+\epsilon}\ln{\dfrac{r_D}{r_S}}\right].
\end{eqnarray}
The above equation is composed of three parts, one accounting for the propagation in Minkowski spacetime, one (depending on $M$) accounting for Schwarzschild and one (depending on $b_0M$) accounting for the effects of the RZ metric.

The second situation is that of gravitational lensing of neutrinos emitted by a distant source and passing in the vicinity of the source of the RZ metric. Therefore we assume that a neutrino produced by some source $S$ travels towards the gravitational lens described by the RZ metric, reaching the point of closest approach $C$ and then proceeds towards the detector $D$ (see Fig.\ref{Fig:weak lensing}). In this case, the integral for the phase in Eq.\eqref{Eq:phase2} can be split into two parts, one between $S$ and $C$ and one between $C$ and $D$. Therefore
\begin{eqnarray} \nonumber 
\Phi_k&=&\dfrac{m_k^2}{2 E_0}\left[ \int_{r_C}^{r_S} \sqrt{\dfrac{\mathcal{A} \mathcal{B}}{\left(1-\dfrac{b^2 \mathcal{A}}{\mathcal{D}}\right)}} dr +\right.\\
 &&+ \left.\int_{r_C}^{r_D} \sqrt{\dfrac{\mathcal{A} \mathcal{B}}{\left(1-\dfrac{b^2 \mathcal{A}}{\mathcal{D}}\right)}} dr\right],
\label{Eq:phase3}
\end{eqnarray}
where $r_S$, $r_D$ and $r_C$ are distances to the source, detector and the closest approach point from the lens, respectively. 
The distance of the point of closest approach $C$ can be
obtained from
\begin{equation}\label{Eq:closest1}
    \left(\dfrac{d r}{d \phi}\right)_0=\dfrac{p_0(r_C) \mathcal{D}}{J_0 \mathcal{B}}=0,
\end{equation}
which in the weak-field approximation gives
\begin{equation}\label{Eq:closest2}
    r_C \approx b\left(1-\dfrac{M}{b}\right).
\end{equation}
Interestingly the point of closest approach does not depend on $b_0$.
Now we integrate Eq.~\eqref{Eq:phase3} substituting Eq.~\eqref{Eq:closest2} and we get 

\begin{eqnarray}\label{Eq:phase4}
\Phi_k &\approx& \dfrac{m_k^2}{2 E_0}\left[\sqrt{r_D^2-b^2}+\sqrt{r_S^2-b^2}+M \left(\dfrac{b}{\sqrt{r_D^2-b^2}}+\right.\right. \nonumber \\
&+&\left.\dfrac{b}{\sqrt{r_S^2-b^2}}+\sqrt{\dfrac{r_D-b}{r_D+b}}+\sqrt{\dfrac{r_S-b}{r_S+b}}\right)+ \nonumber \\ 
&+&\left.\dfrac{2 b_0 M}{ (1+\epsilon)} \ln{\dfrac{r_S r_D}{M^2}}\right].
\end{eqnarray}
We can further simplify the above expression by assuming $b<<r_{S,D}$ and expanding the terms under square roots to obtain
\begin{eqnarray}\label{Eq:phase5}
  \Phi_k &\approx& \dfrac{m_k^2}{2 E_0}(r_S+r_D)\left[1-\dfrac{b^2}{2 r_S r_D}+\dfrac{2 M}{r_S+r_D}+\right. \nonumber \\
  &+& \left. \dfrac{2 b_0 M/(1+\epsilon)}{r_S+r_D} \ln{\dfrac{r_S r_D}{M^2}}\right],
\end{eqnarray}
where we kept only terms up to order $\mathcal{O}\left(b^2/r_{S,D}\right)$. 
Again we can see the distinct contributions due to propagation in flat space, Schwarzschild~\cite{Fornengo_1997} and the RZ metric.

\section{Neutrino oscillation probabilities}\label{sec-5}

\begin{figure*}
    \centering
    \includegraphics[width=5.5cm]{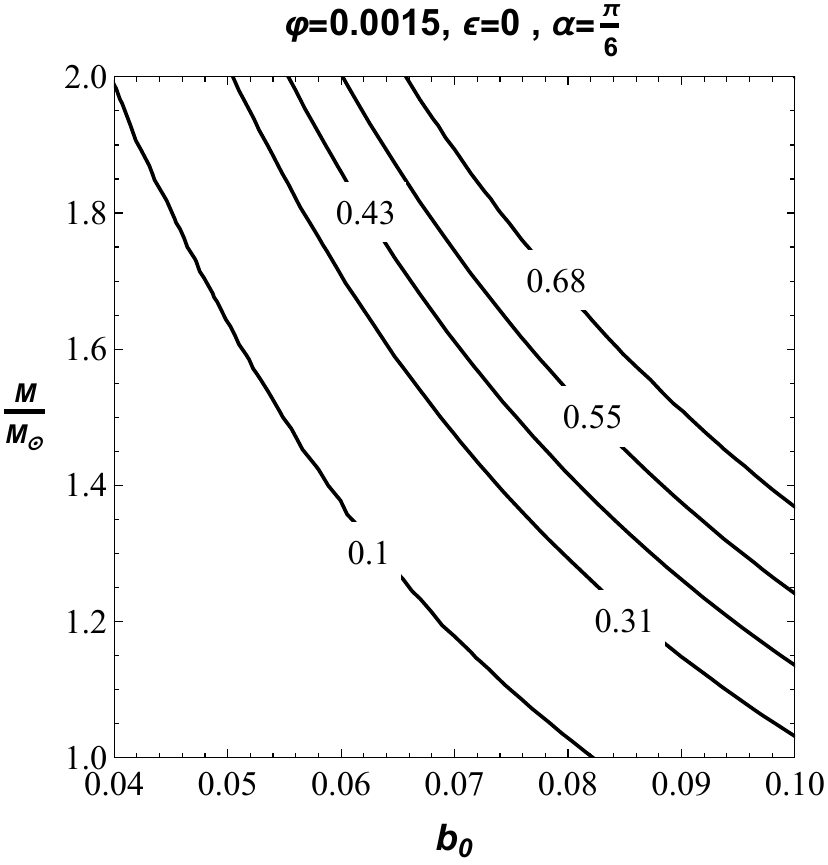}
    \includegraphics[width=5.5cm]{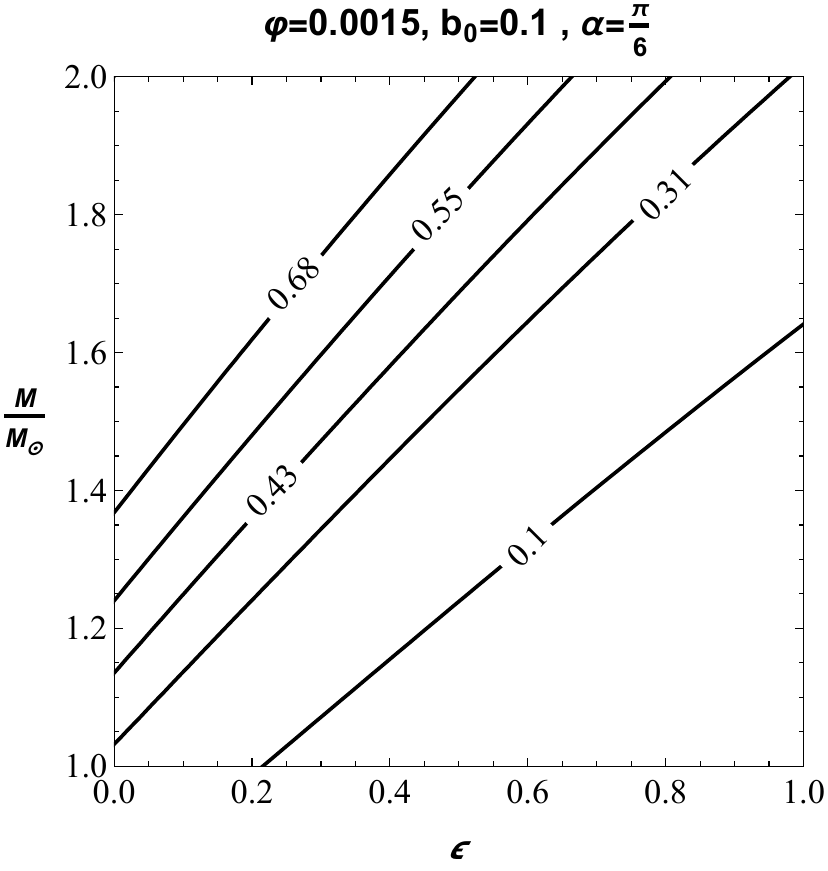}
    \includegraphics[width=5.5cm]{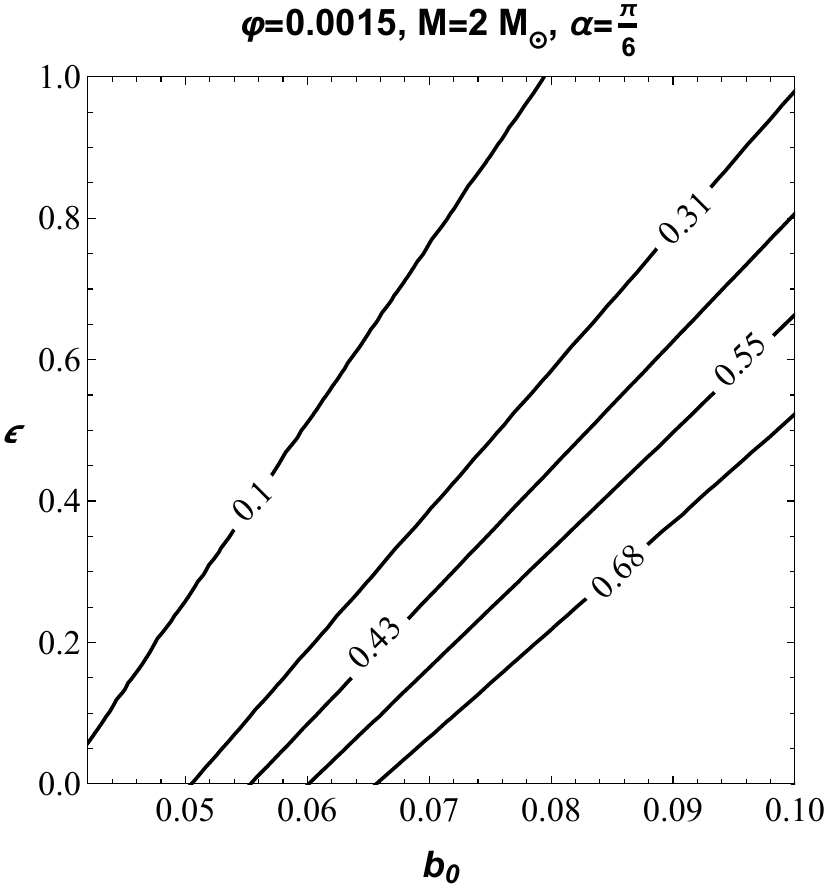}
    \caption{Left panel: The degeneracy between the determination of the mass parameter $M$ and the RZ parameter $b_0$ with $\epsilon = 0$ for a given probability $\mathcal{P}_{e \mu}$ at a fixed angle $\phi$ is illustrated by the $2 D$ contour plot of the implicit function $M(b_0)$ obtained from $\mathcal{P}_{e \mu}(M, b_0) = const$. 
    Middle panel: The degeneracy between the mass parameter $M$ and the RZ parameter $\epsilon$ with $b_0 = 0.1$ for a given probability $\mathcal{P}_{e \mu}$ at a fixed angle $\phi$ is depicted by the $2D$ contour plot of the implicit function $M(\epsilon)$ derived from the equation $\mathcal{P}_{e \mu}(M, \epsilon) = const$. 
    Right panel: The $2D$ contour plot of the implicit function $\epsilon(b_0)$ obtained from $\mathcal{P}_{e \mu}(\epsilon,b_0) = const$ illustrates the degeneracy between the parameter $\epsilon$ and the parameter $b_0$ of the RZ metric for a given probability $\mathcal{P}_{e \mu}$ at a fixed angle $\phi$. Here, the mass parameter $M$ is assumed to be $2M_{\odot}$. Each curve corresponds to a fixed value of $\mathcal{P}_{e \mu}$ as shown on the line itself.}
    \label{fig:degeneracy}
\end{figure*}

Having obtained the phase of oscillation for a gravitationally lensed neutrino, we would now like to calculate the oscillation probabilities of the same. We consider neutrinos with mass eigenstate  $\nu_i$ travelling in the RZ spacetime. The main idea is that a neutrino emitted from the source $S$ can travel along two different paths, say $p$ and $q$, the proper distances of which are different and produce quantum interference at the detector $D$. A neutrino produced in a flavor eigenstate  $|\nu_{\alpha},S\rangle=\cos{\theta}|\nu_1 \rangle+\sin{\theta}|\nu_2\rangle$ at the source S, evolves into 
\begin{equation}
    |\nu_{\alpha},D \rangle = N \sum_i U^{*}_{\alpha i} \sum_p \exp{(-i\Phi^p_i)|\nu_i\rangle},
\end{equation}
where $N$ is a normalization constant and $\Phi^p_i$ is given by Eq.~\eqref{Eq:phase4} with the impact parameter $b_p$ that must be understood as dependent on the path $p$. The oscillation probability $\mathcal{P}$ for neutrino flavor–changing from $\nu_\alpha$ to $\nu_\beta$ at the detector can be written as
\begin{align}\label{Eq:probability1}
 \mathcal{P}_{\alpha\beta}&=|\langle\nu_{\beta}|\nu_\alpha,D \rangle|^2 = \nonumber \\
  & =  |N|^2 \sum_{i,j}  U_{\beta i} U^{*}_{\beta j} U_{\alpha j}U^{*}_{\alpha i}\sum_{p,q}\exp{(-i \Delta \Phi^{pq}_{ij})},
\end{align}
where
\begin{equation}
    |N|^2=\left(\sum_i |U_{\alpha i}|^2 \sum_{p,q} \exp(-i \Delta \Phi^{pq}_{ii})\right)^{-1},
\end{equation}
is the normalization factor \footnote{The same formula in \cite{Chakrabarty:2021bpr} and \cite{Chakrabarty:2023kld} contains typos.} and $\Delta \Phi^{pq}_{ij}$ is the phase difference given by
\begin{equation}\label{Eq:phasedif}
    \Phi^{pq}_{ij}=\Phi_i^p-\Phi_j^q=\Delta m_{ij}^2 A_{pq}+\Delta b_{pq}^2 B_{i j},
\end{equation}
where
\begin{eqnarray}\label{Eq:AB}
   A_{pq}&=&\dfrac{r_S+r_D}{2 E_0}\left[1+\dfrac{2 M}{r_S+r_D}\right.+\nonumber \\
  &+&\left.\dfrac{2 b_0 M}{(r_S+r_D)(1+\epsilon)}\ln{\dfrac{r_S r_D}{M^2}}-\dfrac{\Sigma b^2_{pq}}{4 r_S r_D}\right], \nonumber \\
  B_{ij}&=&-\dfrac{\Sigma m^2_{ij}}{8 E_0}\left(\dfrac{1}{r_S}+\dfrac{1}{r_D}\right).
\end{eqnarray}
In the above equations, the quantities $\Delta m^2_{ij}$, $\Sigma m^2_{ij}$, $\Delta b^2_{pq}$ and $\Sigma b^2_{pq}$ are defined as follows 
\begin{eqnarray}
    && \Delta m^2_{ij}=m^2_{i}-m^2_{j},\hspace{0.5 cm} \Sigma m^2_{ij}=m^2_{i}+m^2_{j}, \nonumber\\
    &&\Delta b^2_{pq}=b^2_{p}-b^2_{q},\hspace{0.75cm} \Sigma b^2_{pq}=b^2_{p}+b^2_{q}.
\end{eqnarray}
We can clearly see that $A_{pq}$ and $B_{ij}$ are symmetric under interchange of their indices, $A_{pq}=A_{qp}$ and $B_{ij}=B_{ji}$. On the other hand, the oscillation probability depends on the sum of individual mass squared of neutrinos $\Sigma m^2_{ij}$ through $\Delta b^2_{pq}$. For those paths for which $\Delta b^2_{pq}$ vanishes, the probability $\mathcal{P}^{lens}_{\alpha \beta}$ is invariant under a shift symmetry $m^2_i \rightarrow m^2_i+C$, while for the paths for which $\Delta b^2_{pq}$ does not vanish, the symmetry $m^2_i \rightarrow m^2_i+C$ is broken. The shift implies $B_{ij} \rightarrow B_{ij}+2C$.

We substitute Eq.~\eqref{Eq:phasedif} in Eq.~\eqref{Eq:probability1} to get
\begin{eqnarray}
     {\mathcal P^{lens}_{\alpha\beta}}&=&|N|^2 \Big[\sum_{i,j}U_{\beta i}U^{*}_{\beta j}U_{\alpha j}U^{*}_{\alpha i}\Big(\sum_{p=q}\exp{(-i\Delta m^2_{ij}A_{pp})}+\nonumber \\
    &+& 2 \sum_{p>q} \cos{(\Delta b^2_{pq} B_{ij})} \exp{(-i \Delta m^2_{ij}}A_{pq})\Big)\Big],
\end{eqnarray}
where 
\begin{equation}
    |N|^2=\left(N_{path}+\sum_i |U_{\alpha i}|^2 \sum_{q>p}2 \cos{(\Delta b^2_{pq}B_{ii})}\right)^{-1}.
\end{equation}
We consider neutrinos traveling in the equatorial plane ($\theta=\pi/2$) for simplicity and in this case, $N_{path}=2$. We can then write the general expression for the probability in the following form
\begin{eqnarray}
    {\mathcal{P}}^{lens}_{\alpha \beta}&=&|N|^2 \left[2 \sum_i |U_{\beta i}|^2|U_{\alpha i}|^2(1+\cos{(\Delta b^2 B_{ii}}))+\right. \nonumber\\
    &+&\sum_{i,j\not=i}U_{\beta i}U^{*}_{\beta j}U_{\alpha j}U^{*}_{\alpha i}\big[\exp{(-i\Delta m^2_{ij} A_{11})}+\nonumber \\
    &+&\exp{(-i\Delta m^2_{ij} A_{22})}+ \nonumber\\
    &+&2 \left.\cos{(\Delta b^2 B_{ij})}\exp{(-i\Delta m^2_{ij}A_{12})} \big]\frac{}{}\right],
\end{eqnarray}
where the normalization factor now is
\begin{equation}
     |N|^2=\left(2+2\sum_i |U_{\alpha i}|^2  \cos{(\Delta b^2_{pq}B_{ii})}\right)^{-1}.
\end{equation}
In order to obtain a quantitative treatment of neutrino lensing in the RZ spacetime, we consider a simple toy model of two neutrino flavors, $\nu_e \rightarrow \nu_{\mu}$. For this transition, the oscillation probability becomes
\begin{eqnarray}\label{eq:prob}
   && {\mathcal{P}}^{lens}_{e \mu}= |N|^2 \sin^2{2 \alpha}\left[\sin^2{\left(\Delta m^2 \dfrac{A_{11}}{2}\right)}+\right.\nonumber \\
   && +\sin^2{\left(\Delta m^2 \dfrac{A_{22}}{2}\right)}-\cos{(\Delta b^2 B_{12})}\cos{(\Delta m^2 A_{12})}+\nonumber \\
   &&+\left.\dfrac{1}{2}\cos{(\Delta b^2 B_{11})}+\dfrac{1}{2}\cos{(\Delta b^2 B_{22})}\right],
\end{eqnarray}
and the normalization constant reduces to
\begin{equation}
        |N|^2 =\frac{1}{2}(1+\cos^2{\alpha}\cos{(\Delta b^2 B_{11})} 
        +\sin^2{\alpha}\cos{(\Delta b^2 B_{22})})^{-1}.
\end{equation}
In this case $\Delta b^2=\Delta b^2_{12}$ and  $\Delta m^2=\Delta m^2_{21}$ while $A_{pq}$ and $B_{ij}$ are given by Eq.~(\ref{Eq:AB}). 
Notice that due to the function $\Delta b^2$ being even, ${\mathcal{P}}^{lens}_{e \mu}$ does not change under the interchange of $b_1$ and $b_2$. However the oscillation probability is sensitive to the neutrino mass ordering, leading to different results for $\Delta m^2>0$ and $\Delta m^2<0$. Additionally, we can see that the oscillation probability is sensitive to slight deviations from Schwarzschild spacetime and spherical symmetry through the involvement of the terms $A_{pq}$ which depend on the RZ deformation parameters $b_0$ and $\epsilon$.

\subsection{Numerical results for the two flavor toy model}

\begin{figure*}[!htb]
    \centering
    \includegraphics[width=5.5cm]{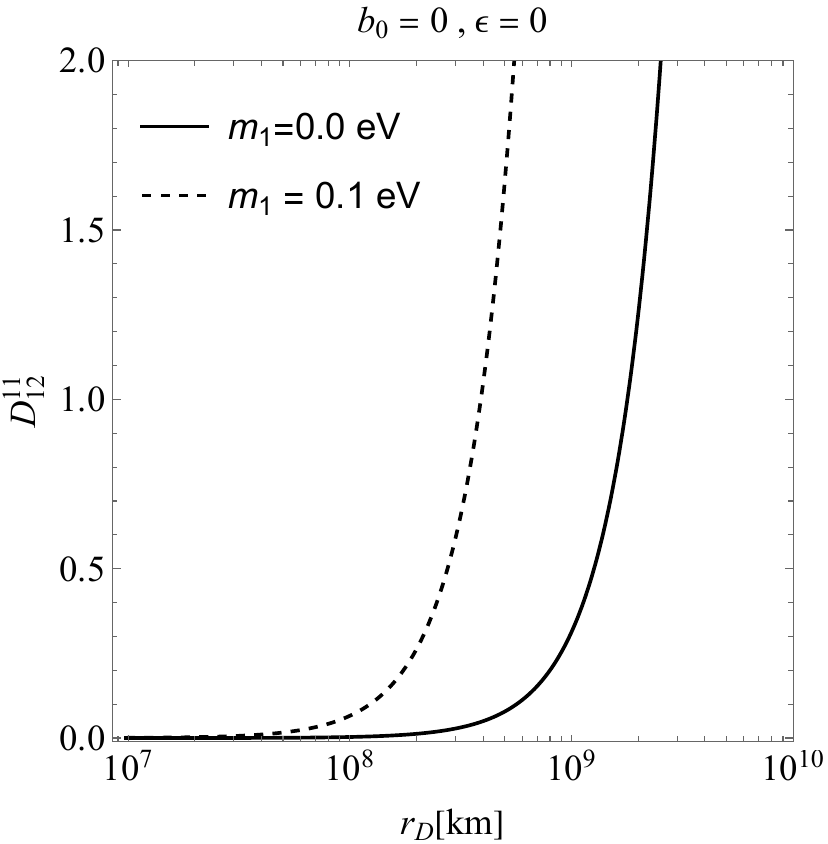}
    \includegraphics[width=5.5cm]{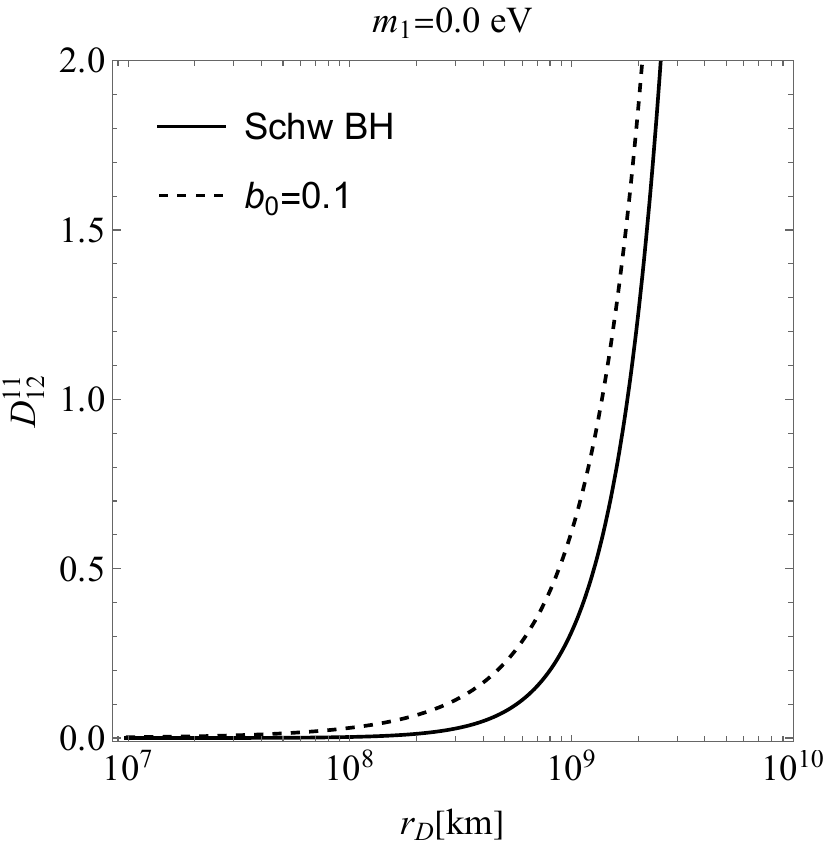}
    \includegraphics[width=5.5cm]{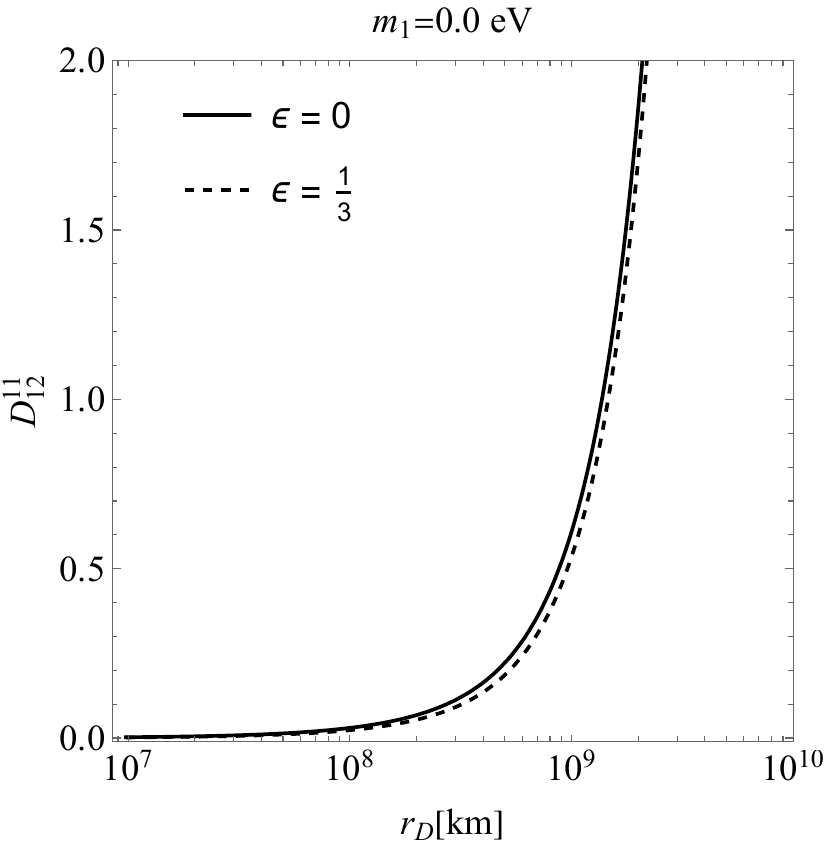}
    \caption{The damping factor $D_{12}^{11}$ as a function of $r_D$ for $\Bar{\sigma}/E_{loc}=10^{-13}$. In the left panel, the solid (dashed) line corresponds to $m_1=0 eV$ ($m_1=0.1 eV$). In the middle panel, the solid and dashed lines correspond to the Schwarzschild and RZ line elements (with $b_0=0.1$ and $\epsilon=0$), respectively.  The right panel is plotted for two different values of the parameter $\epsilon$. Here, $m_1=0.0 eV$ and $b_0=0.1$. The other parameters are $r_S=10^5  r_D$, $\Delta m_{21}^2=m_2^2-m_1^2=10^{-3} eV^2$ and $E_{loc}=10  MeV$.}
    \label{fig:decoherence}
\end{figure*}

For a better understanding, we would like to see
how the probability is affected by the variation of the lensing
parameters in some realistic scenario. Therefore we need to express the impact parameter in terms of known geometrical quantities. 
To this aim we can refer to Fig.\ref{Fig:weak lensing}, which illustrates a schematic diagram of weak lensing in the RZ spacetime. Neutrinos are produced at the point $S$ and are lensed by a massive object whose geometry is described by the RZ spacetime. Eventually neutrinos are detected at the point $D$. From Fig.\ref{Fig:weak lensing} we see that we can express the distances from the massive object of the source and detector in a Cartesian plane $\{x,y\}$ as $r_S(x,y)$ and $r_D(x,y)$, respectively. Alternatively we can consider another coordinate system $\{x',y'\}$ obtained by rotating the original
system $(x,y)$ by an angle $\varphi$ such that $x'=x \cos{\varphi}+y \sin{\varphi}$ and $y'=-x \sin{\varphi}+y \cos{\varphi}$. Let us consider the deflection angle $\delta$ in the rotated frame as
\begin{equation}
    \delta \sim \dfrac{y'_D-b}{x'_D}=-\dfrac{4 M}{b}=-\dfrac{2 R_x}{b},
\end{equation}
where $R_x=2 M=r_0(1+\epsilon)$ and $(x'_D,y'_D)$ is location of the detector. Using the identity $\sin{\varphi}=b/r_S$, the last equation can be written as
\begin{equation}\label{Eq:impactp}
    (2 R_x x_D+b y_D)\sqrt{1-\dfrac{b^2}{r_S^2}}=b^2\left(\dfrac{x_D}{r_S}+1\right)-\dfrac{2 R_x b y_D}{r_S}.
\end{equation}

The solution of Eq.(\ref{Eq:impactp}) gives the impact parameters in terms
of $r_S$, $R_x$ and and the detector's location $(x_D,y_D)$. As an example we shall consider the Sun-Earth system with typical values of the geometrical quantities and assume that the gravitational field of the Sun may be represented by the RZ metric while the Earth's location, $r_D = 10^8 km$, is taken as the detector. We then assume that the source situated behind the Sun at a larger distance $r_S = 10^5r_D$ and it emits relativistic neutrinos with typical energy $E_0 = 10 \ MeV$.  Now assuming a circular trajectory of the detector around the Sun such that $x_D=r_D \cos{\varphi}$ and $y_D=r_D \sin{\varphi}$, we can numerically solve the quartic polynomial given by Eq.~(\ref{Eq:impactp}) and obtain two positive real roots $b_1$ and $b_2$ for every $\varphi$. In this
numerical exercise, the neutrino oscillation probability is calculated only for those value of $b_p$ for which $R_x \ll b_p \ll r_D$. In other words, neutrinos pass far enough from the RZ matter source to consider weak lensing while the detector's location is taken much farther than the impact parameter. The other relevant parameters are $M=1 M_{\odot}$ and $|\Delta m^2|=10^{-3} eV^2$. Note that, these numbers are for illustrative purposes only and in a realistic scenario proper numerical values of the geometric parameters of the model must be considered. 

The oscillation probabilities for the two-flavor toy model of neutrinos are shown in Figs.~\ref{fig2} and~\ref{fig3}. Note that our main aim is to investigate the dependence of the oscillation probability on the parameters of the RZ spacetime. Fig.~\ref{fig2} illustrates the neutrino oscillation probability $\nu_e \rightarrow \nu_{\mu}$ for $b_0=0$ (solid line), which corresponds to Schwarzschild, $b_0=0.01$ (dotted line) and $b_0=0.1$ (dashed line). The figure shows normal hierarchy, i.e. $\Delta m^2 > 0$, in the top panel and inverted hierarchy, i.e. $\Delta m^2 < 0$, in the bottom panel. One can see that the probability depends on the value of the RZ deformation parameter $b_0$. Here, the mixing angle equals to $\alpha=\pi/6$. In Fig.~\ref{fig3} we illustrate the oscillation probability as a function of azimuthal angle $\varphi$. Again it is evident how the parameter $b_0$ affects the oscillation probability while the effect of $\epsilon$ become visible only for $b_0$ sufficiently large.

It is interesting to notice that there is a degeneracy for different values of the RZ parameters which produce the same oscillation probability.
In Fig.~\ref{fig:degeneracy} we plotted the functions $M(b_0)$ (left panel) and $M(\epsilon)$ (middle panel) obtained implicitly from Eq.\eqref{epsilon} for which we obtain the same probability ${\cal P}_{\mu \nu}$ at a given value of the angle $\varphi$. The right panel of Fig.~\ref{fig:degeneracy} shows the values of $\epsilon$ and $b_0$ which give the same probability for a given lens mass $M=2M_{\odot}$ and at a given angle $\varphi$.

\section{Decoherence}\label{sec-6}

\begin{figure*}
    \centering
    \includegraphics{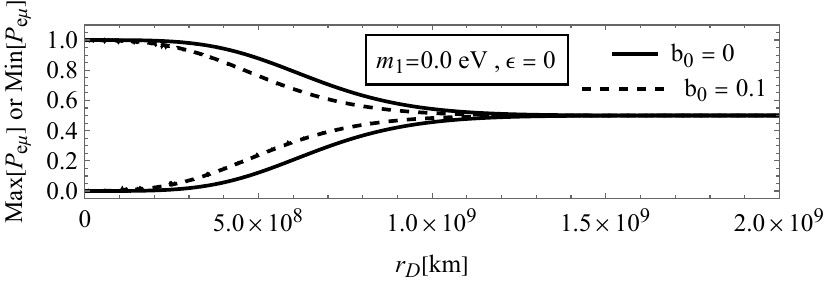}
    \includegraphics{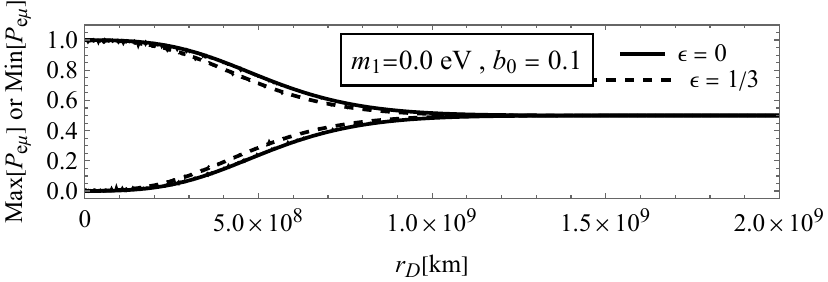}
    \caption{Maximum and minimum transition probability envelop as a function of $r_D$ for two flavor case. The mixing angle is $\alpha=\pi/4$ and all the other parameters are as given in the caption of Fig.~\ref{fig:decoherence}.}
    \label{fig:transitionprob}
\end{figure*}

In this section, we shall shift our attention to the impact of decoherence on the oscillations of neutrinos that are lensed by a gravitational object situated between the source and detector. Instead of the plane-wave approximation that was adopted in the previous analysis, here our argument relies on the assumption that the neutrinos are characterized by Gaussian wave packets. The decoherence length, which is derived from the exponential suppression of flavor transition amplitude, is influenced by the proper time of the geodesic that connects the production and detection events, particularly in a general gravitational setting. In a weak gravitational field, the proper time between two events with a given proper distance is smaller as compared to flat spacetime. Therefore, when a compact object is present, the neutrino wave packets must travel a larger physical distance in space to experience the same amount of proper time before decoherence occurs.

Expressing the probability in wave packet terms is given by (for a detailed discussion, see \cite{Swami_2021})
\begin{equation}\label{prob}
    \mathcal{P}_{\alpha\beta} = \frac{\sum_{i,j} U_{\beta i}^* U_{\alpha i} U_{\beta j}^* U_{\alpha j} \sum_{p,q} e^{-i \Phi_{i j}^{p q}}   e^{- X_{i j}^{p q}}}{\sum_{i} U_{\alpha i} U_{\alpha i}^* \sum_{p,q}e^{-i \Phi_{ii}^{p q}} \ e^{- X_{ii}^{p q}}} ,
\end{equation}
where 
\begin{align}\label{dPhase1}
    \Phi_{ij}^{pq}= \left(\Phi_{i}^p-\Phi_{j}^q\right)- \frac{\Bar{\sigma}^2}{\sigma_D^2 }\left(\Vec{p}^{D}-\Vec{p}^{S}\right) \left(\Vec{\textbf{X}}_i^p-\Vec{\textbf{X}}_j^q\right) \ ,
\end{align}
and
\begin{align}\label{dPhase2}
     \textbf{X}_{ij}^{pq}= \frac{1}{2} \Bar{\sigma}^2 \left(|\Vec{\textbf{X}}_i^p|^2+|\Vec{\textbf{X}}_j^q|^2\right). 
\end{align}
Here $\Bar{\sigma}^2=\sigma_D^2 \sigma_S^2/\left(\sigma_D^2+\sigma_S^2\right)$ with $\sigma_{D,S}$ denoting the standard deviations of the momentum distribution functions at the detector and the source and $\Vec{\textbf{X}}_i^p = \partial_{\Vec{p}}\Phi_{i}^p $. By using the definitions above, it is easy to determine the extent of both oscillations and decoherence in neutrinos that are characterized by Gaussian wave packets and travelling through a weak gravitational field.

In section~\ref{sec-4}, we have detailed the geometric setup that applies also to this case and the computation of the phase $\Phi_i^p$, which in the weak-field regime can be given by Eq.\eqref{Eq:phase5}.
Now, a straightforward evaluation of $\Vec{X}_i^p$ yields

\begin{eqnarray}
       |\Vec{X}_i^p|^2 &\simeq& -\dfrac{m_i^4}{4 E_{loc}^4  {\cal A}(r_S)}(r_S+r_D)^2\left(1-\dfrac{b_p^2}{2 r_S r_D}+\right. \nonumber \\
       & +&\left. \dfrac{2 M}{r_S+r_D} + \dfrac{2 b_0 M/(1+\epsilon)}{r_S+r_D} \ln{\dfrac{r_S r_D}{M^2}}\right)^2 \simeq \nonumber \\
       & \simeq& -\dfrac{m_i^4}{4 E_{loc}^4  {\cal A}(r_S)}(r_S+r_D)^2\left(1-\dfrac{b_p^2}{2 r_S r_D}+\right. \nonumber \\
       & +&\left. \dfrac{4 M}{r_S+r_D} + \dfrac{4b_0 M/(1+\epsilon)}{r_S+r_D} \ln{\dfrac{r_S r_D}{M^2}}\right),
\end{eqnarray}
where $E_{loc}$ represents the energy of neutrinos (in the equal energy approximation) as observed by a local observer at the source, which can be written as~\cite{Swami_2021}
\begin{equation}
    E_{loc}=E_{loc}(r_S)=\frac{E_0}{\sqrt{-{\cal A}(r_S)}}.
\end{equation} 
To measure the decoherence, we will use the effective damping factor $D_{ij}^{pq}$ which is given by 
\begin{align}\label{damping}
    D_{ij}^{pq}=\textbf{X}_{ij}^{pq}-\textbf{X}_{\hat{i}\hat{i}}^{\hat{p}\hat{q}} \ ,
    \end{align}
with $\textbf{X}_{\hat{i}\hat{i}}^{\hat{p}\hat{q}}$ being the smallest among $\textbf{X}_{ij}^{pq}$ and the corresponding $ U_{\alpha \hat{i}} \neq 0 $ and consider the scenario where the source and detector are positioned on opposite sides of the gravitating object. In this case, there are two classical trajectories along which neutrinos can travel, say $p=1$ and $q=2$, distinguished by their impact factors $b_1$ and $b_2$. By aligning the x-axis with the line connecting the neutrino source and RZ matter source, we can select impact parameters such for which $b_1\leq b_2$ for $y\geq0$, and then arrange the neutrino masses in ascending order such that $m_1<m_2<...<m_n$. Using Eq.(\ref{damping}), an appropriate damping factor for the RZ metric can be determined for $y\geq0$ as
\begin{eqnarray}
    D_{ij}^{pq}&=&\textbf{X}_{ij}^{pq}-\textbf{X}_{11}^{11} \simeq -\frac{\Bar{\sigma}^2 \left( r_S+r_D \right)^2 }{8 E_{loc}^4  {\cal A}(r_s)} \left(1+\dfrac{4 M}{r_S+r_D}+\right. \nonumber\\
    &+&\left. \dfrac{4 b_0 M/(1+\epsilon)}{r_S+r_D} \ln{\dfrac{r_S r_D}{M^2}}\right)\times \left[m_i^4\left(1-\frac{b_p^2}{r_S r_D} \right)+\right.\nonumber\\
    &+& \left.m_j^4\left(1-\frac{b_q^2}{r_S r_D} \right)-2m_1^4\left(1-\frac{b_1^2}{r_S r_D} \right)\right]\ .
\end{eqnarray}
There are two ways in which decoherence can occur: (a) due to a mass difference between the lightest and second lightest neutrino mass eigenstate, where $i$ or $j$ is not equal to 1, and (b) due to path difference even when $i=j=1$. However, the second effect is negligible as it arises at a sub-leading order. It is important to note that the contribution from (b) actually decreases in our region of interest, as $b_1<b_2$, and the resulting modifications are extremely small.

Now we can estimate the decoherence length for the example considered above of a Sun-Earth based lensing system (see \cite{Swami_2020}). We consider $R_x=3$ km, $E_{loc}=10 \ MeV$ and $r_S=10^5 r_D$. For simplicity, we consider the co-linear case in which the source of neutrinos, the detector and the gravitating body lie on the same line~\cite{Swami_2021}. The estimation is done for two different values of the lightest neutrino mass $m_1$ while keeping $\Delta m_{21}^2=m_2^2-m_1^2=10^{-3} \ eV^2$ fixed. The obtained results are illustrated in Fig.~\ref{fig:decoherence} where the damping factor is plotted as a function of $r_D$. In the left panel, the solid (dashed) line corresponds to $m_1=0 \ eV$ ($m_1=0.1 \ eV$) for the Schwarzschild spacetime, thus confirming the results of \cite{Swami_2021}. The panel in the middle compares the Schwarzschild black hole (solid line) and the RZ metric with $b_0 = 0.1$ and $\epsilon = 0$ (dashed line) for $m_1=0 \ eV$. The right panel shows the damping factor for the RZ metric with the deformation parameters $b_0 = 0.1, \epsilon = 0$ (solid line) and $b_0 = 0.1, \epsilon = 1/3$ (dashed line). Again we see that the effect of the parameter $b_0$ dominates over that of $\epsilon$.

One can plot the maximum and minimum transition probability envelop as a function of $r_D$ for different values of the $b_0$ and $\epsilon$. This is shown in Fig.~\ref{fig:transitionprob}. To generate the transition probability envelop, we choose the probability's highest and lowest values in the range of $\Delta r_D$ near certain values of $r_D$. For this plot, we have considered the range of $r_D$ from $10^8$ to $2\times10^9 \ {\rm km}$. Also, we choose $\Delta r_D=2\times10^6 \ {\rm km}$ and the mixing angle $\alpha=\pi/4$. The plot consists of $2000$ data points. As we can see from the plot, beyond a certain length, neutrinos lose coherence and the decoherence length has insignificant effects from the deformation parameters of the RZ metric as compared to the absolute neutrino masses which was shown to be very significant in \cite{Swami_2021}.

\section{Summary}\label{sec-7}

Measuring the geometry in the exterior of extreme astrophysical compact objects is one of the goals of several ongoing and future experiments aimed at testing possible deviations from the black hole solutions of classical general relativity.
The RZ metric is a simple parametrized extension of the Schwarzschild geometry that can be used to this aim. 
Proposed measurements that may provide data on the geometry usually rely on the motion of `test particles', the appearance of accretion disks, lensing of photons and potentially lensing of neutrinos. 
In the present article we considered oscillations of neutrino flavors propagating in the RZ spacetime and determined how the detection of neutrinos may in principle be used to constrain the geometry in which they travelled. We showed that the probability of neutrino oscillations depends on the values of the parameters in the RZ metric which characterize the deviation of the geometry from Schwarzschild.
Of course at present such kinds of experiments are still beyond our reach.
However, the model, while presenting an idealized scenario, proves that in principle it may be possible to determine features of the geometry from the observations of neutrinos.
More sophisticated models, such as for example with the inclusion of angular momentum \cite{Swami:2022xet}, are necessary to better characterize the feasibility of such observations.

\section*{Acknowledgement}
DM and HC acknowledge funding support by Nazarbayev University under Faculty-development Competitive Research Grant program for 2022-2024, grant No. 11022021FD2926. M.A. wishes to acknowledge Nazarbayev University for the warm hospitality during his stay in Astana, Kazakhstan.

\bibliographystyle{apsrev4-2}
\bibliography{ref}

\end{document}